\documentclass[a4paper]{jpconf}
\usepackage{graphicx}
\begin{document}

\title{%
Spin-orbital state induced by strong spin-orbit coupling}

\author{Hiroaki Onishi}

\address{%
Advanced Science Research Center,
Japan Atomic Energy Agency,
Tokai, Ibaraki 319-1195, Japan}

%\ead{}

%%%%%%%%%%%%%%%%%%%%%%%%%%%%%%%%%%%%%%%%%%%%%%%%%%%%%%%%%%%%%%%%%%%%%%
% Abstract
%%%%%%%%%%%%%%%%%%%%%%%%%%%%%%%%%%%%%%%%%%%%%%%%%%%%%%%%%%%%%%%%%%%%%%
\begin{abstract}

To clarify a crucial role of a spin-orbit coupling
in the emergence of novel spin-orbital states
in $5d$-electron compounds such as Sr$_{2}$IrO$_{4}$,
we investigate ground state properties of a $t_{\rm 2g}$-orbital Hubbard model
on a square lattice by Lanczos diagonalization.
In the absence of the spin-orbit coupling,
the ground state is spin singlet.
When the spin-orbit coupling is strong enough,
the ground state turns into a weak ferromagnetic state.
The weak ferromagnetic state is a singlet state
in terms of an effective total angular momentum.
Regarding the orbital state,
we find the so-called complex orbital state,
in which real $xy$, $yz$, and $zx$ orbital states are mixed with complex coefficients.

\end{abstract}

%%%%%%%%%%%%%%%%%%%%%%%%%%%%%%%%%%%%%%%%%%%%%%%%%%%%%%%%%%%%%%%%%%%%%%
% Introduction
%%%%%%%%%%%%%%%%%%%%%%%%%%%%%%%%%%%%%%%%%%%%%%%%%%%%%%%%%%%%%%%%%%%%%%
\section{Introduction}

Novel quantum phenomenon driven by
the interplay of strong spin-orbit coupling and Coulomb interaction
has been a subject of significant study in the fields of
condensed matter physics and materials science.
Since spin and orbital degrees of freedom are entangled at a local level
by the spin-orbit coupling,
the total angular momentum provides a good description of
ordered phases and fluctuation properties.
On the other hand, we would envisage possible realization of devices
with exotic functionalities that utilize cross correlation involving
magnetic and electric degrees of freedom,
in which magnetic behavior can be manipulated by an applied electric field,
while electric behavior can be controlled by a magnetic field.

Recently, a $5d$ transition metal oxide Sr$_{2}$IrO$_{4}$,
in which Ir$^{4+}$ ions have five electrons
with a low-spin $(t_{\rm 2g})^{5}$ state,
has attracted growing attention
as a candidate for a novel Mott insulator
characterized by an effective total angular momentum $J_{\rm eff}$=$\frac{1}{2}$,
where ${\bf J}_{\rm eff}$=$-$${\bf L}$+${\bf S}$,
rather than spin commonly observed in $3d$ Mott insulators
\cite{Kim2008,Kim2009}.
The spin-orbit coupling in the $t_{\rm 2g}$ manifold stabilizes the $J_{\rm eff}$=$\frac{1}{2}$ state.
The emergent behavior of the $J_{\rm eff}$=$\frac{1}{2}$ Mott insulator
has been explored in experiments such as
angle resolved photoemission spectroscopy
\cite{Kim2008},
optical conductivity
\cite{Kim2008,Moon2009},
resonant x-ray scattering
\cite{Kim2009},
and
resonant inelastic x-ray scattering
\cite{Ishii2011}.
An isostructural iridate Ba$_{2}$IrO$_{4}$ has also been reported
as the $J_{\rm eff}$=$\frac{1}{2}$ Mott insulator
\cite{Okabe2011}.
Theoretical efforts have been devoted to clarify the characteristics
of the $J_{\rm eff}$=$\frac{1}{2}$ Mott insulator on the basis of
first-principles calculations
\cite{Kim2008,Jin2009}
and model calculations
\cite{Jackeli2009,Watanabe2010,Onishi2011,Wang2011}.

The purpose of this paper is to gain an insight into the spin-orbital state
under the spin-orbit coupling from a microscopic viewpoint.
We investigate ground-state properties of a $t_{\rm 2g}$-orbital Hubbard model
including the spin-orbit coupling by exploiting numerical techniques.
We discuss a spin-orbit-induced ground-state transition
from a spin singlet state to a singlet state in terms of the effective total angular momentum.

%%%%%%%%%%%%%%%%%%%%%%%%%%%%%%%%%%%%%%%%%%%%%%%%%%%%%%%%%%%%%%%%%%%%%%
% Model and numerical method
%%%%%%%%%%%%%%%%%%%%%%%%%%%%%%%%%%%%%%%%%%%%%%%%%%%%%%%%%%%%%%%%%%%%%%
\section{Model and numerical method}

We consider triply degenerate $t_{\rm 2g}$ orbitals on a square lattice
in the $xy$ plane,
and the electron number per site is five,
corresponding to the low-spin state of Ir$^{4+}$ ions.
The $t_{\rm 2g}$-orbital Hubbard model is given by
\begin{eqnarray}
\label{eq:ham}
  H
  &=&
  \sum_{{\bf i},{\bf a},\tau,\tau',\sigma}
  t_{\tau\tau'}^{\bf a}
  (d_{{\bf i}\tau\sigma}^{\dag} d_{{\bf i}+{\bf a}\tau'\sigma}+{\rm h.c.})
  +\lambda\sum_{{\bf i}} {\bf L}_{{\bf i}} \cdot {\bf S}_{{\bf i}}
  +U \sum_{{\bf i},\tau}
  \rho_{{\bf i}\tau\uparrow} \rho_{{\bf i}\tau\downarrow}
  +\frac{U'}{2} \sum_{{\bf i},\sigma,\sigma',\tau\neq\tau'} 
  \rho_{{\bf i}\tau\sigma} \rho_{{\bf i}\tau'\sigma'}
\nonumber\\
  &&
  +\frac{J}{2} \sum_{{\bf i},\sigma,\sigma',\tau\neq\tau'} 
  d_{{\bf i}\tau\sigma}^{\dag} d_{{\bf i}\tau'\sigma'}^{\dag}
  d_{{\bf i}\tau\sigma'} d_{{\bf i}\tau'\sigma}
  +\frac{J'}{2} \sum_{{\bf i},\sigma\neq\sigma',\tau\neq\tau'} 
  d_{{\bf i}\tau\sigma}^{\dag} d_{{\bf i}\tau\sigma'}^{\dag}
  d_{{\bf i}\tau'\sigma'} d_{{\bf i}\tau'\sigma},
\end{eqnarray}
where $d_{{\bf i}\tau\sigma}$ is an annihilation operator for an electron
with spin $\sigma$ (=$\uparrow,\downarrow$)
in orbital $\tau$ (=$xy,yz,zx$)
at site ${\bf i}$,
$\rho_{{\bf i}\tau\sigma}$=$d_{{\bf i}\tau\sigma}^{\dag}d_{{\bf i}\tau\sigma}$,
and ${\bf S}_{\bf i}$ and ${\bf L}_{\bf i}$ represent
spin and orbital angular momentum operators, respectively.
The hopping amplitude is given by
$t_{xy,xy}^{\bf x}$=$t_{zx,zx}^{\bf x}$=$t_{xy,xy}^{\bf y}$=$t_{yz,yz}^{\bf y}$=$t$,
and zero for other combinations of orbitals.
Hereafter, $t$ is taken as the energy unit.
$\lambda$ denotes the spin-orbit coupling.
$U$, $U'$, $J$, and $J$' are
intra-orbital Coulomb,
inter-orbital Coulomb,
exchange (Hund's rule coupling),
and pair-hopping interactions, respectively.
We assume that
$U$=$U'$+$J$+$J'$ due to the rotational symmetry in the local orbital space,
and $J$=$J'$ due to the reality of the $xy$, $yz$, and $zx$ orbital functions.

We numerically investigate ground-state properties
of the $2$$\times$$2$ four-site system
by Lanczos diagonalization.
Because of the three orbitals, the number of bases per site is $4^{3}$=$64$,
and the matrix dimension of the Hamiltonian becomes huge as the system size increases.
In general, the matrix dimension is reduced by decomposing the Hilbert space
into a block-diagonal form by using symmetries of the Hamiltonian.
In the present case,
the total number of electrons is a good quantum number.
We cannot use the $z$ component of the total spin as a good quantum number,
since the spin SU(2) symmetry is broken by the spin-orbit coupling.
We have not utilized the lattice symmetry such as the translational symmetry.
Thus the matrix dimension is $10,626$ for four sites,
while it grows to
%$1,947,792$ for six sites
$377,348,994$ for eight sites.

%%%%%%%%%%%%%%%%%%%%%%%%%%%%%%%%%%%%%%%%%%%%%%%%%%%%%%%%%%%%%%%%%%%%%%
% Results
%%%%%%%%%%%%%%%%%%%%%%%%%%%%%%%%%%%%%%%%%%%%%%%%%%%%%%%%%%%%%%%%%%%%%%
\section{Results}

%%%%% Fig. 1 %%%%%%%%%%%%%%%%%%%%%%%%%%%%%%%%%%%%%
\begin{figure}[t]
\begin{center}
\includegraphics[width=0.9\textwidth]{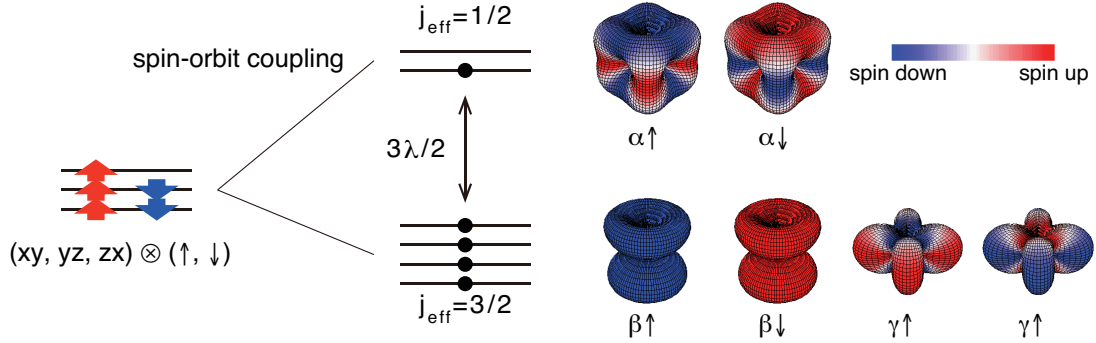}
\end{center}
\caption{
Local electron configuration and wavefunction
of the $t_{\rm 2g}$ orbitals under the spin-orbit coupling.
To visualize the wavefunction,
we draw the surface defined by
$r$=$\sqrt{\sum_{\sigma}\vert\psi(\theta,\phi,\sigma)\vert^{2}}$
in the polar coordinate with $\sigma$ being real spin,
while the color denotes the weight of spin up and down states.
}
\end{figure}
%%%%%%%%%%%%%%%%%%%%%%%%%%%%%%%%%%%%%%%%%%%%%%%%%%

First we briefly discuss the local electron configuration in the atomic limit.
As shown in Fig.~1,
the $t_{\rm 2g}$ level is split by the spin-orbit coupling
into a $j_{\rm eff}$=$\frac{1}{2}$ doublet and a $j_{\rm eff}$=$\frac{3}{2}$ quartet,
in which the eigenstates are characterized by the effective total angular momentum $j_{\rm eff}$.
The eigenstates are
$\vert\alpha\pm\rangle$=%
$\frac{1}{\sqrt{3}}\vert xy \pm\rangle \pm\frac{1}{\sqrt{3}}\vert yz \mp\rangle $$+$$\frac{{\rm i}}{\sqrt{3}}\vert zx \mp\rangle$
with eigenenergy $\lambda$ for the $j_{\rm eff}$=$\frac{1}{2}$ doublet,
and
$\vert\beta\pm\rangle$=%
$\frac{1}{\sqrt{2}}\vert yz \mp\rangle \mp\frac{{\rm i}}{\sqrt{2}}\vert zx \mp\rangle$
and
$\vert\gamma\pm\rangle$=%
$\sqrt{\frac{2}{3}}\vert xy \pm\rangle \mp\frac{1}{\sqrt{6}}\vert yz \mp\rangle $$-$$\frac{{\rm i}}{\sqrt{6}}\vert zx \mp\rangle$
with eigenenergy $-$$\frac{\lambda}{2}$ for the $j_{\rm eff}$=$\frac{3}{2}$ quartet,
where we introduce $\beta$ and $\gamma$ orbitals
to distinguish two Kramers doublets in the $j_{\rm eff}$=$\frac{3}{2}$ quartet,
and pseudospins to label two states in each Kramers doublet.
Note that due to the entanglement of spin and orbital states,
the wavefunction exhibits anisotropic charge and spin distributions.
Since we have five electrons for an Ir$^{4+}$ ion,
the lower $j_{\rm eff}$=$\frac{3}{2}$ quartet is fully occupied,
while the upper $j_{\rm eff}$=$\frac{1}{2}$ doublet is half-filled.

%%%%% Fig. 2 %%%%%%%%%%%%%%%%%%%%%%%%%%%%%%%%%%%%%
\begin{figure}[t]
\begin{center}
\includegraphics[width=0.75\textwidth]{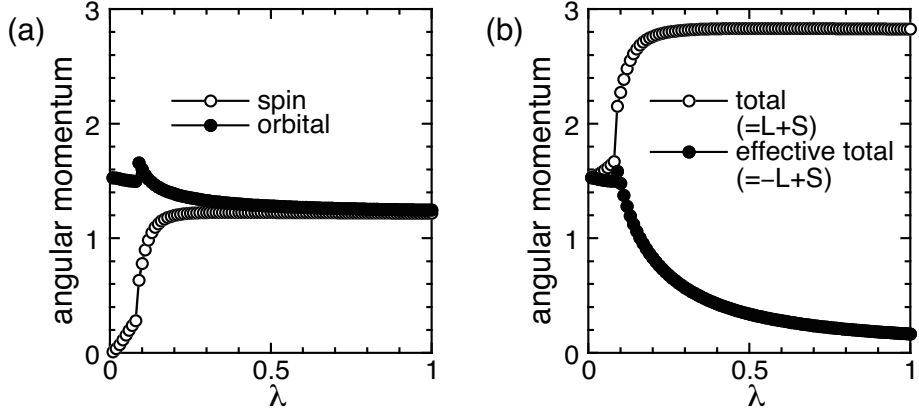}
\end{center}
\caption{
The magnitude of the sum total of angular momenta in the whole system
at $U'$=$10$ and $J$=$2$.
(a) Spin and orbital angular momenta.
(b) Total and effective total angular momenta.
}
\end{figure}
%%%%%%%%%%%%%%%%%%%%%%%%%%%%%%%%%%%%%%%%%%%%%%%%%%

Now we move on to the Lanczos results.
In Fig.~2(a), we present
the magnitude of sum total of spin and orbital angular momenta in the whole system,
defined by
$\langle(\sum_{\bf i}{\bf S}_{\bf i} )^{2}\rangle$=$S_{\rm tot}(S_{\rm tot}+1)$
and
$\langle(\sum_{\bf i}{\bf L}_{\bf i})^{2}\rangle$=$L_{\rm tot}(L_{\rm tot}+1)$,
respectively.
At $\lambda$=$0$,
$S_{\rm tot}$ is found to be zero,
indicating a spin singlet ground state.
As $\lambda$ increases,
$S_{\rm tot}$ gradually increases
and a sudden change occurs at a transition point,
above which the ground state turns to a weak ferromagnetic state.
Note that the induced spin moment is reduced from the maximum value
$\frac{1}{2}$$\times$$4$=$2$.
We also find an abrupt change for $L_{\rm tot}$,
implying that the orbital configuration is reorganized.
Figure~2(b) represents
the magnitude of sum total of total and effective total angular momenta in the whole system,
defined by
$\langle(\sum_{\bf i}{\bf J}_{\bf i} )^{2}\rangle$=$J_{\rm tot}(J_{\rm tot}+1)$
and
$\langle(\sum_{\bf i}{\bf J}_{{\rm eff},{\bf i}})^{2}\rangle$=$J_{\rm eff,tot}(J_{\rm eff,tot}+1)$,
respectively.
$J_{\rm tot}$ is enhanced in the weak ferromagnetic state.
In contrast,
$J_{\rm eff,tot}$ decreases and approaches zero in the limit of large $\lambda$.
Thus the weak ferromagnetic state is regarded as a singlet state
in terms of the effective total angular momentum.

To clarify how the orbital state changes as $\lambda$ varies,
we measure the charge density
in each of the $t_{\rm 2g}$ orbitals,
$n_{\tau}$=$\sum_{\sigma}\langle\rho_{{\bf i}\tau\sigma}\rangle$,
in different two basis sets.
Note that due to the translational invariance,
the charge density is equivalent in all sites.
In Fig.~3(a), we plot the charge density in the real $xy$, $yz$, and $zx$ orbitals.
At $\lambda$=$0$,
we find that $n_{xy}$=1.5 and $n_{yz}$=$n_{zx}$=1.75,
since the itinerancy of orbitals depends on the hopping direction
and holes preferably occupy the $xy$ orbitals rather than the $yz$ and $zx$ orbitals.
As $\lambda$ increases,
$n_{xy}$, $n_{yz}$, and $n_{zx}$ approach $\frac{5}{3}$,
indicating that $xy$, $yz$, and $zx$ orbital states are mixed with equal weight.
Note that $n_{yz}$ and $n_{zx}$ are equivalent irrespective of $\lambda$.
Transforming the basis set into the complex $\alpha$, $\beta$, and $\gamma$ orbitals,
we can see the characteristics of the orbital state
from the viewpoint of the effective total angular momentum.
As shown in Fig.~3(b),
we observe a sharp change at the transition point,
implying again the reorganization of the orbital configuration.
For large $\lambda$,
the $\alpha$ orbitals are singly occupied and relevant to the low-energy property,
while the $\beta$ and $\gamma$ orbitals are fully occupied,
consistent with the local electron configuration in the atomic limit,
as shown in Fig.~1.
This clearly indicates the emergence of a complex orbital state,
in which spin and orbital states are entangled with complex number coefficients.

%%%%% Fig. 3 %%%%%%%%%%%%%%%%%%%%%%%%%%%%%%%%%%%%%
\begin{figure}[t]
\begin{center}
\includegraphics[width=0.75\textwidth]{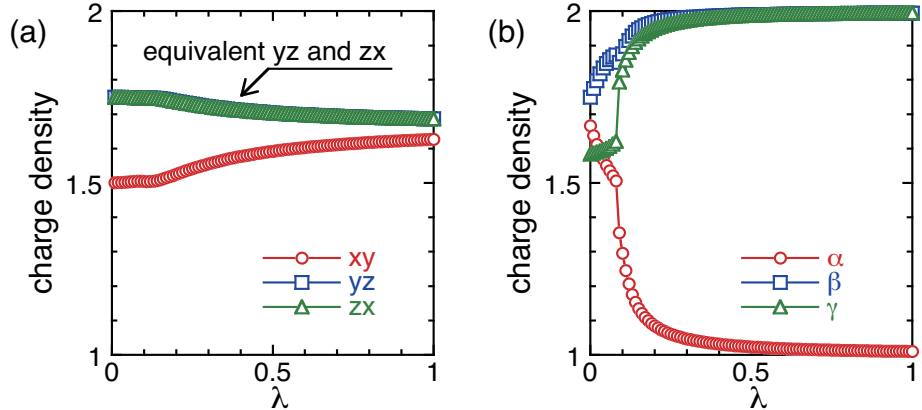}
\end{center}
\caption{
The charge density in each of the $t_{\rm 2g}$ orbitals
for different basis sets
at $U'$=$10$ and $J$=$2$.
(a) Real $xy$, $yz$, and $zx$ orbitals.
(b) Complex $\alpha$, $\beta$, and $\gamma$ orbitals.
}
\end{figure}
%%%%%%%%%%%%%%%%%%%%%%%%%%%%%%%%%%%%%%%%%%%%%%%%%%

%%%%%%%%%%%%%%%%%%%%%%%%%%%%%%%%%%%%%%%%%%%%%%%%%%%%%%%%%%%%%%%%%%%%%%
% Summary
%%%%%%%%%%%%%%%%%%%%%%%%%%%%%%%%%%%%%%%%%%%%%%%%%%%%%%%%%%%%%%%%%%%%%%
\section{Summary}

We have studied ground-state properties of the $t_{\rm 2g}$-orbital Hubbard model
with the spin-orbit coupling by Lanczos diagonalization.
We have found that due to the spin-orbit coupling,
the ground state changes
from the spin singlet
to the singlet in terms of the effective total angular momentum.
The complex orbital state characterized by $J_{\rm eff}$=$\frac{1}{2}$ emerges
in the spin-orbit-induced state.
To clarify novel magnetism in $5d$ transision metal oxides such as Sr$_{2}$IrO$_{4}$,
an important issue is to understand the excitation dynamics under the strong spin-orbit coupling,
which we will discuss elsewhere in future.

%%%%%%%%%%%%%%%%%%%%%%%%%%%%%%%%%%%%%%%%%%%%%%%%%%%%%%%%%%%%%%%%%%%%%%
% Acknowledgments
%%%%%%%%%%%%%%%%%%%%%%%%%%%%%%%%%%%%%%%%%%%%%%%%%%%%%%%%%%%%%%%%%%%%%%
\ack

The author thanks
G. Khaliullin, S. Maekawa, M. Mori,
T. Shirakawa, H. Watanabe, and S. Yunoki
for discussions.
Part of numerical calculations were performed
on the supercomputer at Japan Atomic Energy Agency.
This work was supported by Grant-in-Aid for Scientific Research of
Ministry of Education, Culture, Sports, Science, and Technology of Japan.

%%%%%%%%%%%%%%%%%%%%%%%%%%%%%%%%%%%%%%%%%%%%%%%%%%%%%%%%%%%%%%%%%%%%%%
% References
%%%%%%%%%%%%%%%%%%%%%%%%%%%%%%%%%%%%%%%%%%%%%%%%%%%%%%%%%%%%%%%%%%%%%%

\end{document}